\documentclass[pra,groupedaddress,showpacs,showkeys]{revtex4}

\newtheorem{theorem}{Theorem}[section]

\newtheorem{corollary}[theorem]{Corollary}
\newtheorem{definition}[theorem]{Definition}

\newcommand{\qed}{\nobreak \ifvmode \relax \else
      \ifdim\lastskip<1.5em \hskip-\lastskip
      \hskip1.5em plus0em minus0.5em \fi \nobreak
      \vrule height0.75em width0.5em depth0.25em\fi}

\begin{document}

\title{Consistency of the Adiabatic Theorem}
\author{M.S. Sarandy, L.-A. Wu, D.A. Lidar}
\affiliation{Chemical Physics Theory Group, Department of Chemistry, and Center for Quantum Information and
Quantum Control, University of Toronto, 80 St.
George Street, Toronto, Ontario, M5S 3H6, Canada}

\begin{abstract}
We review the quantum adiabatic approximation for
closed systems, and its recently introduced generalization to open
systems (M.S. Sarandy and D.A. Lidar, e-print quant-ph/0404147). We also
critically examine a recent argument claiming that there is an
inconsistency in the adiabatic theorem for closed quantum systems
[K.P. Marzlin and B.C. Sanders, Phys. Rev. Lett. 93, 160408 (2004)] and point out how an
incorrect manipulation of the adiabatic theorem may lead one to obtain
such an inconsistent result.
\end{abstract}



\maketitle


\section{Introduction}


The adiabatic theorem~\cite{Ehrenfest:16,Born:28,Kato:50,Messiah:book} is
one of the oldest and most widely used general tools in quantum mechanics.
The theorem concerns the evolution of systems subject to slowly varying
Hamiltonians. Roughly, its content is that if a state is an instantaneous
eigenstate of a sufficiently slowly varying $H$ at one time then it will
remain an eigenstate at later times, while its eigenenergy evolves
continuously. When the slowness assumption is relaxed transitions become
weakly allowed \cite{Berry:87,Nakagawa:87,Sun:88,Wu:89}. The role of the
adiabatic theorem in the study of slowly varying quantum mechanical systems
spans a vast array of fields and applications, such as the Landau-Zener
theory of energy level crossings in molecules \cite{Landau:32,Zener:32},
quantum field theory \cite{Gellmann:51}, and Berry's phase \cite{Berry:84}.
In recent years geometric phases \cite{Wilczek:84} have been proposed to
perform quantum information processing~\cite%
{ZanardiRasseti:99,ZanardiRasseti:2000,Ekert-Nature}, with adiabaticity
assumed in a number of schemes for geometric quantum computation (e.g.,~\cite%
{Pachos:00,Duan-Science:01,Pachos:02,Fazio:03}). Additional interest in
adiabatic processes has arisen in connection with the concept of adiabatic
quantum computing, in which slowly varying Hamiltonians appear as a
promising mechanism for the design of new quantum algorithms and even as an
alternative to the conventional quantum circuit model of quantum computation~%
\cite{Farhi:00,Farhi:01,Aharonov:04}.

More recently, in Ref.~\cite{SarandyLidar:04}, the adiabatic theorem was
generalized to the case of open quantum systems, i.e., quantum systems
coupled to an external environment. Instead of making use of eigenstates of
the Hamiltonian, adiabaticity is defined through the Jordan canonical form
of the generator of the master equation governing the dynamics of the
system. This new framework allowed for the derivation of an adiabatic
approximation which includes the case of systems evolving in the presence of
noise. This issue is particularly important in the context of quantum
information processing, where environment induced decoherence is viewed as a
fundamental obstacle on the path to the construction of quantum computers
(e.g.,~\cite{LidarWhaley:03}).

The aim of this paper is to review the adiabatic approximation in quantum
mechanics for both closed and open quantum systems as well as to point out
how an incorrect manipulation of the adiabatic theorem can yield an
inconsistent result. Indeed, in a recent paper entitled \textquotedblleft
Inconsistency in the application of the adiabatic theorem\textquotedblright\ 
\cite{Marzlin:04} the authors argue that there may be an inconsistency in
the adiabatic theorem for closed quantum systems. We show here how this
inconsistency can be resolved. Related discussions can be found in Refs.~%
\cite{Tong:04,Pati:04,Wu:04}.


\section{The quantum adiabatic approximation for closed systems}


\label{closed}


\subsection{Condition on the Hamiltonian}


Let us begin by reviewing the adiabatic approximation in closed quantum
systems, which evolve unitarily through a time-dependent Schr\"{o}dinger
equation 
\begin{equation}
H(t)\,|\psi (t)\rangle =i\,|{\dot{\psi}}(t)\rangle ,  \label{se}
\end{equation}%
where $H(t)$ denotes the Hamiltonian and $|\psi (t)\rangle $ is a quantum
state in a $D$-dimensional Hilbert space. We use units where ${\hbar }=1$.
For simplicity we assume that the spectrum of $H(t)$ is entirely discrete
and nondegenerate. Thus we can define an instantaneous basis of
eigenenergies by 
\begin{equation}
H(t)\,|n(t)\rangle =E_{n}(t)\,|n(t)\rangle ,  \label{ebh}
\end{equation}%
with the set of eigenvectors {$|n(t)\rangle $} chosen to be orthonormal. In
this simplest case, where to each energy level there corresponds a unique
eigenstate we can \emph{define adiabaticity as the regime associated to an
independent evolution of the instantaneous eigenvectors of} $H(t)$.

This means that instantaneous eigenstates at one time evolve continuously to
the corresponding eigenstates at later times, and that their corresponding
eigenenergies do not cross. In particular, if the system begins its
evolution in a particular eigenstate $|n(0)\rangle $ then it will evolve to
the instantaneous eigenstate $|n(t)\rangle $ at a later time $t$, without
any transition to other energy levels.

It is conceptually useful to point out that the relationship between slowly
varying Hamiltonians and adiabatic behavior can be demonstrated directly
from a simple manipulation of the Schr\"{o}dinger equation: recall that $%
H(t) $ can be diagonalized by a unitary similarity transformation 
\begin{equation}
H_{d}(t)=U^{-1}(t)\,H(t)\,U(t),  \label{hdc}
\end{equation}%
where $H_{d}(t)$ denotes the diagonalized Hamiltonian and $U(t)$ is a
unitary transformation. Multiplying Eq.~(\ref{se}) by $U^{-1}(t)$ and using
Eq.~(\ref{hdc}) we obtain 
\begin{equation}
H_{d}\,|\psi \rangle _{d}=i\,|{\dot{\psi}}\rangle _{d}-i\,{\dot{U}}%
^{-1}|\psi \rangle ,  \label{sed}
\end{equation}%
where $|\psi \rangle _{d}\equiv U^{-1}|\psi \rangle $ is the state of the
system in the basis of eigenvectors of $H(t)$. Upon considering that $H(t)$
changes slowly in time, i.e. $dH(t)/dt\approx 0$, we may also assume that
the unitary transformation $U(t)$ and its inverse $U^{-1}(t)$ are slowly
varying operators, yielding 
\begin{equation}
H_{d}(t)\,|\psi (t)\rangle _{d}=i\,|{\dot{\psi}}(t)\rangle _{d}.
\label{eq:Had}
\end{equation}%
Thus, since $H_{d}(t)$ is diagonal, the system evolves separately in each
energy sector, ensuring the validity of the adiabatic approximation.

Now let 
\begin{equation}
g_{nk}(t)\equiv E_{n}(t)-E_{k}(t)  \label{eq:g}
\end{equation}%
be the energy gap between level $n$ and $k$ and let $T$ be the total
evolution time. One may then state \emph{a general validity condition for
adiabatic behavior} as follows: 
\begin{equation}
\max_{0\leq t\leq T}\left\vert \frac{\langle k|{\dot{H}}|n\rangle }{g_{nk}}%
\right\vert \,\ll \,\min_{0\leq t\leq T}\left\vert {g_{nk}}\right\vert .
\label{vcc}
\end{equation}%
Note that the left-hand side of Eq.~(\ref{vcc}) has dimensions of frequency
and hence must compared to the relevant physical frequency scale, which can
be proved to be given by the gap $g_{nk}$~\cite%
{Messiah:book,Mostafazadeh:book}. [In fact, Eq.~(\ref{vcc}) will be seen to
be a direct consequence of the adiabatic condition derived in Subsection~\ref%
{condtimeclosed}.] The interpretation of the adiabaticity condition~(\ref%
{vcc}) is that for all pairs of energy levels, the expectation value of the
time-rate-of-change of the Hamiltonian, in units of the gap, must be small
compared to the gap. For a discussion of the adiabatic regime when
there is no gap in the energy spectrum see Ref.~\cite{Avron:98}.

In order to obtain Eq.~(\ref{vcc}), let us expand $|\psi (t)\rangle $ in terms of the basis of
instantaneous eigenvectors of $H(t)$: 
\begin{equation}
|\psi (t)\rangle =\sum_{n=1}^{D}a_{n}(t)\,e^{-i\int_{0}^{t}dt^{\prime
}E_{n}(t^{\prime })}\,|n(t)\rangle ,  \label{ep}
\end{equation}%
with $a_{n}(t)$ being complex functions of time. Substitution of Eq.~(\ref%
{ep}) into Eq.~(\ref{se}) and multiplying the result by $\langle k(t)|$, we
have 
\begin{equation}
{\dot{a}}_{k}=-\sum_{n}a_{n}\langle k|{\dot{n}}\rangle
\,e^{-i\int_{0}^{t}dt^{\prime }g_{nk}(t^{\prime })}.  \label{an2}
\end{equation}%
A useful expression for $\langle k|{\dot{n}}\rangle $, for $k\neq n$, can be
found by taking a time derivative of Eq.~(\ref{ebh}) and multiplying the
resulting expression by $\langle k|$, which reads 
\begin{equation}
\langle k|{\dot{n}}\rangle =\frac{\langle k|{\dot{H}}|n\rangle }{g_{nk}}%
\quad (n\neq k).  \label{knee}
\end{equation}%
Therefore Eq.~(\ref{an2}) can be written as 
\begin{equation}
{\dot{a}}_{k}=-a_{k}\langle k|{\dot{k}}\rangle -\sum_{n\neq k}a_{n}\frac{%
\langle k|{\dot{H}}|n\rangle }{g_{nk}}\,e^{-i\int_{0}^{t}dt^{\prime
}g_{nk}(t^{\prime })}.  \label{anf}
\end{equation}%
Adiabatic evolution is ensured if the coefficients $a_{k}(t)$ evolve
independently from each other, i.e., if their dynamical equations do not
couple. As is apparent from Eq.~(\ref{anf}), this requirement is fulfilled
when the condition (\ref{vcc}) is imposed.

In the case of a degenerate spectrum of $H(t)$, Eq.~(\ref{knee}) holds only
for eigenstates $|k\rangle $ and $|n\rangle $ for which $E_{n}\neq E_{k}$.
Taking into account this modification in Eq.~(\ref{anf}), it is not
difficult to see that the adiabatic approximation generalizes to the
statement that each degenerate eigenspace of $H(t)$, instead of individual
eigenvectors, has independent evolution, whose validity conditions given by
Eq.~(\ref{vcc}) are to be considered over eigenvectors with distinct
energies. Thus, in general one can define adiabatic dynamics of closed
quantum systems as follows:

\begin{definition}
\label{defc} A closed quantum system is said to undergo adiabatic dynamics
if its Hilbert space can be decomposed into decoupled Schr\"{o}%
dinger-eigenspaces with distinct, time-continuous, and non-crossing
instantaneous eigenvalues of $H(t)$.
\end{definition}


\subsection{Condition on the total evolution time}

\label{condtimeclosed} 

A very useful alternative is to express the adiabaticity condition in terms
of the total evolution time $T$. We shall consider for simplicity a
nondegenerate $H(t)$; the generalization to the degenerate case is also
possible. Taking the initial state as the eigenvector $|m(0)\rangle $, with $%
a_{m}(0)=1$, the condition for adiabatic evolution can be stated as follows:%
\begin{equation}
T\gg \frac{\mathcal{F}}{\mathcal{G}^{2}},  \label{timead2}
\end{equation}%
where 
\begin{equation}
\mathcal{F}=\max_{0\leq s\leq 1}|\langle k(s)|\frac{dH(s)}{ds}|m(s)\rangle
|\,,\,\,\,\,\,\,\,\,\,\,\mathcal{G}=\min_{0\leq s\leq 1}|g_{mk}(s)|\,.
\end{equation}%
Eq.~(\ref{timead2}) can be interpreted as stating the total evolution time
must be much larger than the norm of the time-derivative of the Hamiltonian
divided by the square of the energy gap. It gives an important validity
condition for the adiabatic approximation, which has been used, e.g., to
determine the running time required by adiabatic quantum algorithms~\cite%
{Farhi:00,Farhi:01,Aharonov:04}. By using the time variable transformation~(%
\ref{nt}), one can show that Eq.~(\ref{timead2}) is indeed equivalent to the
adiabatic condition~(\ref{vcc}) on the Hamiltonian.

To derive Eq.~(\ref{timead2}), let us rewrite Eq.~(\ref{anf}) as follows~%
\cite{Gottfried:book}: 
\begin{equation}
e^{i\gamma _{k}(t)}\,\frac{\partial }{\partial t}\left(
a_{k}(t)\,e^{-i\gamma _{k}(t)}\right) =-\sum_{n\neq k}a_{n}\frac{\langle k|{%
\dot{H}}|n\rangle }{g_{nk}}\,e^{-i\int_{0}^{t}dt^{\prime }g_{nk}(t^{\prime
})},  \label{adtti}
\end{equation}%
where $\gamma _{k}(t)$ denotes the Berry's phase \cite{Berry:84} associated
to the state $|k\rangle $: 
\begin{equation}
\gamma _{k}(t)=i\int_{0}^{t}dt^{\prime }\langle k(t^{\prime })|{\dot{k}}%
(t^{\prime })\rangle .
\end{equation}%
Now let us define a normalized time $s$ through the variable transformation 
\begin{equation}
t=sT,\,\,\,\,\,0\leq s\leq 1.  \label{nt}
\end{equation}%
Then, by performing the change $t\rightarrow s$ in Eq.~(\ref{adtti}) and
integrating we obtain 
\begin{equation}
a_{k}(s)\,e^{-i\gamma _{k}(s)}=a_{k}(0)-\sum_{n\neq k}\int_{0}^{s}ds^{\prime
}\frac{F_{nk}(s^{\prime })}{g_{nk}(s^{\prime })}e^{-iT\int_{0}^{s^{\prime
}}ds^{\prime \prime }g_{nk}(s^{\prime \prime })},  \label{akint}
\end{equation}%
where 
\begin{equation}
F_{nk}(s)=a_{n}(s)\,\langle k(s)|\frac{dH(s)}{ds}|n(s)\rangle \,e^{-i\gamma
_{k}(s)}.
\end{equation}%
However, for an adiabatic evolution as defined above, the coefficients $%
a_{n}(s)$ evolve without any mixing, which means that $a_{n}(s)\approx
a_{n}(0)\,e^{i\gamma _{n}(s)}$. Therefore 
\begin{equation}
F_{nk}(s)=a_{n}(0)\,\langle k(s)|\frac{dH(s)}{ds}|n(s)\rangle \,e^{-i(\gamma
_{k}(s)-\gamma _{n}(s))}.
\end{equation}%
In order to arrive at a condition on $T$ it is useful to separate out the
fast oscillatory part from Eq.~(\ref{akint}). Thus, the integrand in Eq.~(%
\ref{akint}) can be rewritten as 
\begin{equation}
\frac{F_{nk}(s^{\prime })}{g_{nk}(s^{\prime })}e^{-iT\int_{0}^{s^{\prime
}}ds^{\prime \prime }g_{nk}(s^{\prime \prime })}=\frac{i}{T}\left[ \frac{d}{%
ds^{\prime }}\left( \frac{F_{nk}(s^{\prime })}{g_{nk}^{2}(s^{\prime })}%
e^{-iT\int_{0}^{s^{\prime }}ds^{\prime \prime }g_{nk}(s^{\prime \prime
})}\right) -\,e^{-iT\int_{0}^{s^{\prime }}ds^{\prime \prime
}g_{nk}(s^{\prime \prime })}\frac{d}{ds^{\prime }}\left( \frac{%
F_{nk}(s^{\prime })}{g_{nk}^{2}(s^{\prime })}\right) \right] .  \label{ricc}
\end{equation}%
Substitution of Eq.~(\ref{ricc}) into Eq.~(\ref{akint}) results in 
\begin{equation}
a_{k}(s)\,e^{-i\gamma _{k}(s)}=a_{k}(0)+\frac{i}{T}\sum_{n\neq k}\left( 
\frac{F_{nk}(0)}{g_{nk}^{2}(0)}-\frac{F_{nk}(s)}{g_{nk}^{2}(s)}%
e^{-iT\int_{0}^{s}ds^{\prime }g_{nk}(s^{\prime })}+\,\int_{0}^{s}ds^{\prime
}\,e^{-iT\int_{0}^{s^{\prime }}ds^{\prime \prime }g_{nk}(s^{\prime \prime })}%
\frac{d}{ds^{\prime }}\frac{F_{nk}(s^{\prime })}{g_{nk}^{2}(s^{\prime })}%
\right) .  \label{akfinal}
\end{equation}%
A condition for the adiabatic regime can be obtained from Eq.~(\ref{akfinal}%
) if the last integral vanishes for large $T$. Let us assume that, as $%
T\rightarrow \infty $, the energy difference remains nonvanishing. We
further assume that $d\{F_{nk}(s^{\prime })/g_{nk}^{2}(s^{\prime
})\}/ds^{\prime }$ is integrable on the interval $\left[ 0,s\right] $. Then
it follows from the Riemann-Lebesgue lemma~\cite{Churchill:book} that the
last integral in Eq.~(\ref{akfinal}) vanishes in the limit $T\rightarrow
\infty $ (due to the fast oscillation of the integrand)~\cite%
{RiemannLebesgue}. What is left are therefore only the first two terms in
the sum over $n\neq k$\ of Eq.~(\ref{akfinal}). Thus, a general estimate of
the time rate at which the adiabatic regime is approached can be expressed
by 
\begin{equation}
T\gg \frac{F}{g^{2}},  \label{timead}
\end{equation}%
where 
\begin{equation}
F=\max_{0\leq s\leq 1}|a_{n}(0)\,\langle k(s)|\frac{dH(s)}{ds}|n(s)\rangle
|,\,\,\,\,\,\,\,\,\,\,g=\min_{0\leq s\leq 1}|g_{nk}(s)|\,,
\end{equation}%
with $\max $ and $\min $ taken over all $k$ and $n$. Eq.~(\ref{timead2}) is
then obtained as the special case when the system starts its evolution in a
particular eigenstate of $H(t)$.


\subsection{Higher-order corrections to the adiabatic approximation}

\label{adiabcorr} 

When the Hamiltonian of a quantum system changes slowly, but not extremely
slowly, the degenerate eigenspaces of $H(t)$ (or individual eigenvectors in
the case of nondegenerate spectrum) will not evolve completely independently
from each other and, therefore, the dynamical equation~(\ref{an2}) will
weakly couple distinct eigenspaces of $H(t)$. Then, for non-extremely slowly
varying Hamiltonians, the adiabatic solution is actually a zeroth-order
approximation and higher-order corrections must be considered. Some
higher-order adiabatic approximation methods have been proposed~\cite%
{Berry:87,Nakagawa:87,Sun:88,Wu:89}. Here we shall review, for the
non-degenerate case, the method proposed by Wu in Ref.~\cite{Wu:89}. Let us
begin by expanding the state vector $|\psi (t)\rangle $ in the instantaneous
eigenbasis, as in Eq.~(\ref{ep}). Then, by using the normalized time $s$
introduced in Eq.~(\ref{nt}) we obtain the following matrix form for the Schr%
\"{o}dinger equation 
\begin{equation}
\frac{d\psi (s)}{ds}=K(s)\psi (s),
\end{equation}%
where $K(s)$ is an anti-Hermitian matrix with elements 
\begin{equation}
K_{mn}(s)=-\left\langle m(s)\right\vert \frac{d}{ds}\left\vert
n(s)\right\rangle \exp \left( iT\int_{0}^{s}ds^{\prime }\,g_{mn}(s^{\prime
})\right) .
\end{equation}%
The matrix $K(s)$ can be separated into a diagonal matrix $D(s)$ and an
off-diagonal matrix $O(s)$, yielding 
\begin{equation}
K(s)=D(s)+O(s).  \label{kdo}
\end{equation}%
The evolution operator $U(s)$ for the system satisfies the equation 
\begin{equation}
\frac{dU(s)}{ds}=K(s)U(s),\,\,\,\,\,\,\text{(}{\text{with}}\,\,\,U(0)=1\text{%
)}  \label{evop}
\end{equation}%
which, after integration and use of Eq.~(\ref{kdo}), becomes 
\begin{equation}
U(s)=1+\int_{0}^{s}ds_{1}\,[D(s_{1})+O(s_{1})]+\int_{0}^{s}ds_{1}%
\int_{0}^{s_{1}}ds_{2}\,[D(s_{1})+O(s_{1})][D(s_{2})+O(s_{2})]+\,\ldots
\end{equation}%
Now let us define 
\begin{equation}
U^{(0)}(s)=1+\int_{0}^{s}ds_{1}\,D(s_{1})+\int_{0}^{s}ds_{1}%
\int_{0}^{s_{1}}ds_{2}\,D(s_{1})D(s_{2})+\,\ldots ,
\end{equation}%
which involves only the diagonal parts. Moreover, for $n>0$, we denote $%
U^{(n)}(s)$ as the sum of all the integrals with $n$ off-diagonal $O(s)$
factors in the integrand. Therefore the evolution operator can be expanded
in powers of the off-diagonal matrices $O(s)$ as $U(s)=\sum_{n=0}U^{(n)}(s)$%
. It can be shown~\cite{Wu:89} that the $n^{\mathrm{th}}$ term $U^{(n)}(s)$
can be expressed through the lower order term $U^{(n-1)}(s)$ by means of the
recurrence equation 
\begin{equation}
U^{(n)}(s)=\int_{0}^{s}ds^{\prime }\,U^{(0)}(s^{\prime })\,O(s^{\prime
})\,U^{(n-1)}(s^{\prime }).
\end{equation}%
The expression above means that, by knowing the zeroth-order evolution
operator $U^{(0)}(s)$, which exactly yields the adiabatic approximation, one
can obtain $U^{(1)}(s)$, and then $U^{(2)}(s)$ and so on. The adiabatic case
corresponds to no transitions, while a correction $U^{(n)}(s)$ of order $%
n\geq 1$ implies the existence of $n$ transitions between different energy
levels \cite{Wu:89}. From this perspective one can interpret the adiabatic
approximation as the zeroth order term in a perturbation theory in the
number of transitions between energy levels connected by the time-varying
Hamiltonian.


\section{The quantum adiabatic approximation for open quantum systems}

\label{open} 


In this section we review our recently introduced generalization of the
adiabatic theorem to the case of open quantum systems \cite{SarandyLidar:04}%
. The motivations for considering such a generalization are many. The most
fundamental is that the concept of a closed system is, of course, an
idealization, and in reality all experimentally accessible systems are open.
Thus applications of the adiabatic theorem for open systems include, among
others, geometric phases (where open system effects have received
considerable recent attention, e.g., Refs. \cite{geomphase-open}), quantum
information processing, and molecular dynamics in condensed phases.

In the following we first introduce notation for open systems, then discuss
the generalized adiabatic theorem.

\subsection{The dynamics of open quantum system}


Consider a quantum system $S$ coupled to an environment, or bath $B$ (with
respective Hilbert spaces $\mathcal{H}_{S},\mathcal{H}_{B}$), evolving
unitarily under the total system-bath Hamiltonian $H_{SB}$. The exact system
dynamics is given by tracing over the bath degrees of freedom \cite%
{Breuer:book} 
\begin{equation}
\rho (t)=\mathrm{Tr}_{B}[U(t)\rho _{SB}(0)U^{\dag }(t)],  \label{system}
\end{equation}%
where $\rho (t)$ is the system state, $\rho _{SB}(0)=\rho (0)\otimes \rho
_{B}(0)$ is the initially uncorrelated system-bath state, and $U(t)=\mathcal{%
T}\mathsf{\exp }(-i\int_{0}^{t}H_{SB}(t^{\prime })dt^{\prime })$ ($\mathcal{T%
}$ denotes time-ordering). Such an evolution is completely positive and
trace preserving \cite{Breuer:book,Kraus:71,Alicki:87}. 
Under certain approximations, it is possible to convert Eq.~(\ref{system})
into the convolutionless form
\begin{eqnarray}
{\dot{\rho}}(t) &=& \mathcal{L}(t) \rho (t).
\label{eq:t-Lind}
\end{eqnarray}
An important example is 
\begin{eqnarray}
{\dot{\rho}}(t) &=& 
-i\left[ H(t),\rho (t) \right] +\frac{1}{2}
\sum_{i=1}^{N}\left([\Gamma _{i}(t),\rho (t) \Gamma^{\dagger }_{i}(t)] 
+[\Gamma_{i}(t)\rho (t), \Gamma^{\dagger }_{i}(t)]\right).
\label{eq:t-Lind2}
\end{eqnarray}%
Here $H(t)$ is the time-dependent effective Hamiltonian of the open system and 
$\Gamma _{i}(t)$ are time-dependent operators describing the system-bath
interaction. In the literature, Eq.~(\ref{eq:t-Lind2}) with time-\emph{in}dependent 
operators $\Gamma _{i}$ is usually referred to as the Markovian dynamical semigroup, 
or Lindblad equation \cite{Breuer:book,Alicki:87,Gorini:76,Lindblad:76} [see also
Ref.~\cite{Lidar:CP01} for a simple derivation of
Eq.~(\ref{eq:t-Lind2}) from Eq.~(\ref{system})]. However,
the case with time-dependent coefficients is also permissible under
certain restrictions \cite{Lendi:86}.
The Lindblad equation requires the
assumption of a Markovian bath with vanishing correlation
time. Equation (\ref{eq:t-Lind}) can be more general; for example, it
applies to the case of non-Markovian convolutionless master equations
studied in Ref.~\cite{Breuer:04}.
Here we will consider the class of convolutionless master
equations (\ref{eq:t-Lind}). In a slight
abuse of nomenclature, we will henceforth refer to the time-dependent
generator $\mathcal{L}(t)$ as the Lindblad superoperator, and the $\Gamma
_{i}(t)$ as Lindblad operators.

Conceptually, the difficulty in the transition of an adiabatic approximation
from closed to open quantum systems is that the notion of Hamiltonian
eigenstates is lost, since the Lindblad superoperator -- the generalization
of the Hamiltonian -- cannot in general be diagonalized. It is then not a
priori clear what should take the place of the adiabatic eigenstates. This
difficulty was solved in Ref.~\cite{SarandyLidar:04} by introducing the idea
that this role is played by \emph{adiabatic Jordan blocks of the Lindblad
superoperator}. The Jordan canonical form \cite{Horn:book}, with its
associated left- and right-eigenvectors, is in this context the natural
generalization of the diagonalization of the Hamiltonian. In this direction,
it is convenient to work in the superoperator formalism, wherein the density
matrix is represented by a $D^{2}$-dimensional \textquotedblleft coherence
vector\textquotedblright\ 
\begin{equation}
|\rho \rangle \rangle =\left( 
\begin{array}{cccc}
\rho _{1} & \rho _{2} & \cdots & \rho _{D^{2}}%
\end{array}%
\right) ^{t},  \label{vcv}
\end{equation}%
and the Lindblad superoperator $\mathcal{L}$ becomes a $D^{2}\times D^{2}$%
-dimensional supermatrix \cite{Alicki:87}. We use the double bracket
notation to indicate that we are not working in the standard Hilbert space
of state vectors. More generally, coherence vectors live in Hilbert-Schmidt
space:\ a state space of linear operators endowed with an inner product that
can be defined, for general vectors $u$ and $v$, as 
\begin{equation}
(u,v)\equiv \langle \langle u|v\rangle \rangle \equiv \frac{1}{{\mathcal{N}}}%
{\text{Tr}}\left( u^{\dagger }v\right) .  \label{ip}
\end{equation}%
where ${\mathcal{N}}$ is a normalization factor. Adjoint elements $\langle
\langle v|$ in the dual state space are given by row vectors defined as the
transpose conjugate of $|v\rangle \rangle $: $\langle \langle
v|=(v_{1}^{\ast },v_{2}^{\ast },...,v_{D^{2}}^{\ast })$. A density matrix
can then be expressed as a discrete superposition of states over a complete
basis in this vector space, with appropriate constraints on the coefficients
so that the requirements of Hermiticity, positive semi-definiteness and unit
trace of $\rho $ are observed. Thus, representing the density operator in
general as a coherence vector, we can rewrite Eq.~(\ref{eq:t-Lind}) in a
superoperator language as 
\begin{equation}
\mathcal{L}(t)\,|\rho (t)\rangle \rangle =|{\dot{\rho}}(t)\rangle \rangle ,
\label{le}
\end{equation}%
where $\mathcal{L}$ is now a supermatrix. This master equation generates
non-unitary evolution, since $\mathcal{L}(t)$ is non-Hermitian and hence
generally non-diagonalizable. However, one can always transform $\mathcal{L}$
into the Jordan canonical form \cite{Horn:book}, where it has a
block-diagonal structure. This is achieved via the similarity transformation 
\begin{equation}
\mathcal{L}_{J}(t)=S^{-1}(t)\,\mathcal{L}(t)\,S(t),  \label{jd}
\end{equation}%
where $\mathcal{L}_{J}(t)=\mathrm{diag}(J_{1},...,J_{m})$ denotes the Jordan
form of $\mathcal{L}(t)$. The Jordan blocks $J_{\alpha }$, of dimension $%
n_{\alpha }$, are always of the form: 
\begin{equation}
J_{\alpha }=\left( 
\begin{array}{ccccc}
\lambda _{\alpha } & 1 & 0 & ... & 0 \\ 
0 & \lambda _{\alpha } & 1 & ... & 0 \\ 
\vdots & \ddots & \ddots & \ddots & \vdots \\ 
0 & ... & 0 & \lambda _{\alpha } & 1 \\ 
0 & ... & ... & 0 & \lambda _{\alpha }%
\end{array}%
\right) .  \label{ljmatg}
\end{equation}%
To each Jordan block are associated a left and a right eigenvector with
eigenvalue $\lambda _{\alpha }$, which can in general be complex. The number 
$m$ of Jordan blocks is given by the number of linearly independent
eigenstates of $\mathcal{L}(t)$, with each eigenstate associated to a
different block $J_{\alpha }$. Since $\mathcal{L}(t)$ is non-Hermitian, we
generally do not have a basis of eigenstates, whence some care is required
in order to find a basis for describing the density operator. It can be
shown~\cite{SarandyLidar:04} that instantaneous right $\left\{ |\mathcal{D}%
_{\beta }^{(j)}(t)\rangle \rangle \right\} $ and left $\left\{ \langle
\langle \mathcal{E}_{\alpha }^{(i)}(t)|\right\} $ bases in the state space
of linear operators can always be systematically constructed, with the
following suitable features:

\noindent $\bullet$ Orthonormality condition: 
\begin{equation}
\langle \langle \mathcal{E}_{\alpha }^{(i)}(t)|\mathcal{D}_{\beta
}^{(j)}(t)\rangle \rangle =\,_{J}\langle \langle \mathcal{E}_{\alpha }^{(i)}|%
\mathcal{D}_{\beta }^{(j)}\rangle \rangle _{J}=\delta _{\alpha \beta }\delta
^{ij}.  \label{lrr}
\end{equation}

\noindent $\bullet $ Invariance of the Jordan blocks under the action of the
Lindblad super-operator: 
\begin{eqnarray}
&&\mathcal{L}(t)\,|\mathcal{D}_{\alpha }^{(j)}(t)\rangle \rangle =|\mathcal{D%
}_{\alpha }^{(j-1)}(t)\rangle \rangle +\lambda _{\alpha }(t)\,|\mathcal{D}%
_{\alpha }^{(j)}(t)\rangle \rangle .  \label{ldo} \\
&&\langle \langle \mathcal{E}_{\alpha }^{(i)}(t)|\,\mathcal{L}(t)=\langle
\langle \mathcal{E}_{\alpha }^{(i+1)}(t)|+\langle \langle \mathcal{E}%
_{\alpha }^{(i)}(t)|\,\lambda _{\alpha }(t).  \label{lro}
\end{eqnarray}%
with $|\mathcal{D}_{\alpha }^{(-1)}\rangle \rangle \equiv 0$ and $\langle
\langle \mathcal{E}_{\alpha }^{(n_{\alpha })}|\equiv 0$. Here the subscripts
enumerate Jordan blocks ($\alpha \in \{1,...,m\}$), while the superscripts
enumerate basis states inside a given Jordan block ($i,j\in
\{0,...,n_{\alpha }-1\}$).


\subsection{Adiabatic conditions for open quantum system}


Before stating explicitly the conditions for adiabatic evolution, we
provide a formal definition of adiabaticity for the case of open systems:

\begin{definition}
\label{def:open-ad} An open quantum system is said to undergo adiabatic
dynamics if its Hilbert-Schmidt space can be decomposed into decoupled
Lindblad--Jordan-eigenspaces with distinct, time-continuous, and
non-crossing instantaneous eigenvalues of $\mathcal{L}(t)$.
\end{definition}

This definition is a natural extension for open systems of the idea of
adiabatic behavior. Indeed, in this case the master equation
(\ref{eq:t-Lind}) can be decomposed into sectors with different and
separately evolving 
Lindblad-Jordan eigenvalues. The more familiar notion of closed-system
adiabaticity is obtained as a special case when the Lindblad superoperator
is Hermitian:\ in that case it can be diagonalized and the Jordan blocks all
become one-dimensional, with corresponding real eigenvalues (that correspond
to energy \emph{differences}). The splitting into Jordan blocks of the
Lindblad superoperator is achieved through the choice of a basis which
preserves the Jordan block structure as, for example, the sets of right $%
\left\{ |\mathcal{D}_{\beta }^{(j)}(t)\rangle \rangle \right\} $ and left $%
\left\{ \langle \langle \mathcal{E}_{\alpha }^{(i)}(t)|\right\} $ vectors
introduced above. Such a basis generalizes the notion of Schr\"{o}%
dinger-eigenvectors. Based on this concept of adiabaticity, we state below
(without the proofs) several theorems which have been derived in Ref.~\cite%
{SarandyLidar:04}.

\begin{theorem}
\label{t1} A sufficient condition for open quantum system adiabatic dynamics
as given in Definition~\ref{def:open-ad} is: 
\begin{equation}
\max_{0\leq s\leq 1}\,\left\vert \sum_{p=1}^{(n_{\alpha }-i)}\left(
\prod_{q=1}^{p}\sum_{k_{q}=0}^{(j-S_{q-1})}\right) \frac{\langle \langle 
\mathcal{E}_{\alpha }^{(i+p-1)}|\frac{d\mathcal{L}}{ds}|\mathcal{D}_{\beta
}^{(j-S_{p})}\rangle \rangle }{(-1)^{S_{p}}\,\omega _{\beta \alpha
}^{p+S_{p}}}\right\vert \ll 1,  \label{vc}
\end{equation}%
where $s=t/T$ is the scaled time and 
\begin{equation}
\omega _{\beta \alpha }(t)=\lambda _{\beta }(t)-\lambda _{\alpha
}(t),\,\,\,\,\,\,\,S_{q}=\sum_{s=1}^{q}k_{s}\,\,\,\,\,\,\,\text{(}{\text{with%
}}\,\,\,S_{0}=0\text{)},\,\,\,\,\left(
\prod_{q=1}^{p}\sum_{k_{q}=0}^{(j-S_{q-1})}\right) \equiv
\sum_{k_{1}=0}^{j-S_{0}}\cdots \sum_{k_{p}=0}^{j-S_{p-1}},
\end{equation}%
and where $\lambda _{\beta }\neq \lambda _{\alpha }$, with $i$ and $j$
denoting arbitrary indices associated to the Jordan blocks $\alpha $ and $%
\beta $, respectively.
\end{theorem}

The role of the energy differences that appear in the equations for the closed case 
is played here by the (in general complex-valued) difference between Jordan eigenvalues 
$\omega _{\beta \alpha }$, while the norm of the time-derivative of the 
Hamiltonian is replaced here by the norm of $\frac{d\mathcal{L}}{ds}$, 
evaluated over and inside Jordan blocks. Eq.~(\ref{vc})
gives a means to estimate the accuracy of the adiabatic approximation via
the computation of the time derivative of the Lindblad superoperator acting
on right and left vectors. The norm used in Eq.~(\ref{vc}) can be simplified
by considering only the term with maximum absolute value, which results in:

\begin{corollary}
\label{c1os} A sufficient condition for open quantum system adiabatic
dynamics is 
\begin{equation}
\mathcal{N}_{ij}^{n_{\alpha }n_{\beta }}\max_{0\leq s\leq 1}\,\left\vert 
\frac{\langle \langle \mathcal{E}_{\alpha }^{(i+p-1)}|\frac{d\mathcal{L}}{ds}%
|\mathcal{D}_{\beta }^{(j-S_{p})}\rangle \rangle }{\omega _{\beta \alpha
}^{p+S_{p}}}\right\vert \ll 1,
\end{equation}%
where the $\max $ is taken for any $\alpha \neq \beta $, and over all
possible values of $i\in \{0,...,n_{\alpha }-1\}$, $j\in \{0,...,n_{\beta
}-1\}$, and $p$, with 
\begin{equation}
\mathcal{N}_{ij}^{n_{\alpha }n_{\beta }}=\frac{(n_{\alpha }-i+1+j)!}{%
(1+j)!(n_{\alpha }-i)!}-1.  \label{numberterms}
\end{equation}
\end{corollary}

The factor $\mathcal{N}_{ij}^{n_{\alpha }n_{\beta }}$ given in Eq.~(\ref%
{numberterms}) is just the number of terms of the sums in Eq.~(\ref{vc}). We
have included a superscript $n_{\beta }$, even though there is no explicit
dependence on $n_{\beta }$, since $j\in \{0,...,n_{\beta }-1\}$.

Furthermore, an adiabatic condition for a slowly varying Lindblad
super-operator can directly be obtained from Eq.~(\ref{vc}), yielding:

\begin{corollary}
A simple sufficient condition for open quantum system adiabatic dynamics is $%
{\dot{\mathcal{L}}}\approx 0$.
\end{corollary}

Note that this condition is in a sense too strong, since it need not be the
case that $\dot{\mathcal{L}}$ is small in general (i.e., for all its matrix
elements). 

Just as in the closed-systems case, one can also express the condition for
adiabaticity in terms of the total time of evolution. To this end, we expand
the density matrix for an arbitrary time $t$ in the instantaneous right
eigenbasis $\left\{ |{\mathcal{D}_{\beta }^{(j)}(t)\rangle \rangle }\right\} 
$ as 
\begin{equation}
|\rho (t)\rangle \rangle =\frac{1}{2}\sum_{\beta =1}^{m}\sum_{j=0}^{n_{\beta
}-1}p_{\beta }^{(j)}(t)\,e^{\int_{0}^{t}\lambda _{\beta }(t^{\prime
})dt^{\prime }}\,|\mathcal{D}_{\beta }^{(j)}(t)\rangle \rangle ,
\label{rtime}
\end{equation}%
where $m$ is the number of Jordan blocks and $n_{\beta }$ is the dimension
of the block $J_{\beta }$. We emphasize that we are assuming that there are
no eigenvalue crossings in the spectrum of the Lindblad superoperator during
the evolution. We also define 
\begin{equation}
V_{\beta \alpha }^{(ijp)}(s)=p_{\beta }^{(j)}(s)\langle \langle \mathcal{E}%
_{\alpha }^{(i+p-1)}(s)|\frac{d\mathcal{L}(s)}{ds}|\mathcal{D}_{\beta
}^{(j-S_{p})}(s)\rangle \rangle   \label{vbapj}
\end{equation}%
and 
\begin{equation}
\Omega _{\beta \alpha }(t)=\int_{0}^{t}\omega _{\beta \alpha }(t^{\prime
})\,dt^{\prime }.  \label{Ombapj}
\end{equation}%
Then the following adiabatic time condition can be established \cite%
{SarandyLidar:04}:

\begin{theorem}
\label{t3} Consider an open quantum system governed by a Lindblad
superoperator $\mathcal{L}(s)$. 
Then the adiabatic dynamics in the interval $0\leq s\leq 1$ occurs if and only if the following 
time conditions, obtained for each coefficient $p_{\alpha}^{(i)}(s)$, are satisfied:
\begin{eqnarray}
T &\gg& \max_{0\leq s\leq 1} \left\vert \,\sum_{\beta \,|\,\lambda _{\beta
}\neq \lambda _{\alpha }}\sum_{j,p}\,(-1)^{S_{p}}
\left[ \frac{V_{\beta \alpha }^{(ijp)}(0)}{\omega _{\beta \alpha
}^{p+S_{p}+1}(0)}-\frac{V_{\beta \alpha }^{(ijp)}(s)\,e^{T\,\Omega _{\beta
\alpha }(s)}}{\omega _{\beta \alpha }^{p+S_{p}+1}(s)}
+\int_{0}^{s}ds^{\prime }\,e^{T\,\Omega _{\beta \alpha
}(s^{\prime })}\frac{d}{ds^{\prime }}\frac{V_{\beta \alpha
}^{(ijp)}(s^{\prime })}{\omega _{\beta \alpha }^{p+S_{p}+1}(s^{\prime })}%
\right] \right\vert .  \label{eq:tadscg}
\end{eqnarray}
\end{theorem}

Theorem \ref{t3} provides a very general condition for adiabaticity in open
quantum systems. Equation (\ref{eq:tadscg}) simplifies in a number of situations.

\begin{itemize}
\item Adiabaticity is guaranteed whenever $V_{\beta \alpha }^{(ijp)}(s)$ 
vanishes for all $\lambda_\alpha \neq \lambda_\beta $. 

\item Adiabaticity is similarly guaranteed whenever $V_{\beta \alpha }^{(ijp)}(s)$,
which can depend on $T$ through $p_{\beta }^{(j)}$, vanishes for all $\lambda_\alpha
,\lambda_\beta $ such that $\mathrm{Re}(\Omega _{\beta \alpha })>0$ and does not
grow faster, as a function of $T$, than $\exp (T|\,{\mathrm{Re}}\Omega
_{\beta \alpha }|)$ for all $\lambda_\alpha ,\lambda_\beta $ such that $\mathrm{Re}(\Omega
_{\beta \alpha })<0$.

\item When $\mathrm{Re}(\Omega _{\beta \alpha })=0$ and $\mathrm{Im}(\Omega
_{\beta \alpha })\neq 0$ the integral in inequality (\ref{eq:tadscg}) vanishes
in the infinite time limit due to the Riemann-Lebesgue lemma~\cite%
{Churchill:book}, as in the closed case discussed before. In this case,
again, adiabaticity is guaranteed provided $p_{\beta }^{(j)}(s)$ [and hence $V_{\beta \alpha }^{(ijp)}(s)$] 
does not diverge as a function of $T$ in the limit $T\rightarrow \infty$.

\item When $\mathrm{Re}(\Omega _{\beta \alpha })>0$, the adiabatic regime
can still be reached for large $T$ provided that $p_{\beta }^{(j)}(s)$ contains a
decaying exponential which compensates for the growing exponential due to $%
\mathrm{Re}(\Omega _{\beta \alpha })$.

\item Even if there is an overall growing exponential in inequality (\ref%
{eq:tadscg}), adiabaticity could take place over a finite time interval $%
[0,T_{\ast }]$ and, afterwards, disappear. In this case, which would be an
exclusive feature of open systems, the crossover time $T_{\ast }$ would be
determined by an inequality of the type $T\gg a+b\exp (cT)$, with $c>0$. The
coefficients $a,b$ and $c$ are functions of the system-bath interaction.
Whether the latter inequality can be solved clearly depends on the values of 
$a,b,c$, so that a conclusion about adiabaticity in this case is model
dependent.
\end{itemize}

A simpler sufficient condition can be derived from Eq. (\ref{eq:tadscg}) by
considering the term with maximum absolute value in the sum. This procedure
leads to the following corollary:

\begin{corollary}
\label{ct3} A sufficient time condition for the adiabatic regime of an open
quantum system governed by a Lindblad superoperator $\mathcal{L}(t)$ is 
\begin{eqnarray}
T &\gg& \mathcal{M}_{ij}^{n_\alpha n_\beta} \, \max_{0\le s\le 1} \left\vert
\, \frac{V_{\beta \alpha }^{(ijp)}(0)}{\omega _{\beta \alpha}^{p+S_{p}+1}(0)}
-\frac{V_{\beta \alpha }^{(ijp)}(s)\,e^{T\,\Omega_{\beta \alpha }(s)}}{%
\omega _{\beta \alpha}^{p+S_{p}+1}(s)} 
+\int_{0}^{s}ds^{\prime }\,e^{T\,\Omega _{\beta \alpha }(s^{\prime
})}\frac{d}{ds^{\prime }} \frac{V_{\beta \alpha }^{(ijp)}(s^{\prime })}{%
\omega _{\beta\alpha }^{p+S_{p}+1}(s^{\prime })} \right\vert,
\label{eq:tadcol}
\end{eqnarray}
where $\max $ is taken over all possible values of the indices $%
\lambda_\alpha \neq \lambda_\beta $, $i$, $j$, and $p$, with 
\begin{eqnarray}
&&\mathcal{M}_{ij}^{n_\alpha n_\beta} = \sum_{\beta \,|\,\lambda _{\beta
}\neq \lambda _{\alpha}} \sum_{j=0}^{(n_{\beta}-1)}\sum_{p=1}^{(n_{\alpha
}-i)}\left( \prod_{q=1}^{p}\sum_{k_{q}=0}^{(j-S_{q-1})}\right) 1  
= \Lambda_{\beta\alpha} \left[ \frac{(n_\alpha+n_\beta-i+1)!}{%
(n_\alpha-i+1)!n_\beta!}-n_\beta-1, \right]  \label{Nlt}
\end{eqnarray}
where $\Lambda_{\beta\alpha}$ denotes the number of Jordan blocks such that $%
\lambda_\alpha \neq \lambda_\beta$.
\end{corollary}

Further discussion of the physical significance of these adiabaticity
conditions, as well as an illustrative example, can be found in Ref.~\cite%
{SarandyLidar:04}. The application of the adiabatic theorem for open quantum
systems to problems in quantum information processing (e.g., in the context
of adiabatic quantum computing \cite{Farhi:00,Farhi:01,Aharonov:04}) and
geometric phases~\cite{Thomaz:03,Carollo:04,Sanders:04}, seems particularly appealing as a venue for future
research.


\section{The Marzlin-Sanders inconsistency}



The adiabatic theorem can be deceptively simple when it is not carefully
interpreted. In a recent paper entitled K.-P. Marzlin and B.C. Sanders argue
that there may be an inconsistency in the adiabatic theorem for closed
quantum systems \cite{Marzlin:04}, when the change in instantaneous
adiabatic eigenstates is significant. Here we simplify their argument and
show where exactly is the fallacy that leads one to conclude that there is
such an inconsistency.

\subsection{The condition for adiabaticity revisited}


Let us consider a quantum system evolving unitarily under the Schr\"{o}%
dinger equation~(\ref{se}). At the initial time $t_{0}$ the system is
assumed to be in the particular instantaneous energy eigenstate $%
|E_{0}(t_{0})\rangle $. For a general time $t$ the evolution of the system
is described by 
\begin{equation}
|\psi (t)\rangle =U(t,t_{0})|E_{0}(t_{0})\rangle ,  \label{ge}
\end{equation}%
where $U(t,t_{0})=\mathcal{T}\exp (-i\int_{t_{0}}^{t}H(t^{\prime
})dt^{\prime })$ is the unitary evolution operator. Assuming that the
Hamiltonian changes slowly in time and that $|E_{0}(t_{0})\rangle $ is
non-degenerate, the adiabatic theorem implies that 
\begin{equation}
|\psi (t)\rangle =e^{-i\int^{t}E_{0}}e^{i\beta _{0}}|E_{0}(t)\rangle 
\label{adp}
\end{equation}%
where $\int^{t}E_{0}\equiv \int_{t_{0}}^{t}E_{0}(t^{\prime })dt^{\prime }$
and the Berry's phase $\beta _{0}$ is given by $\beta _{0}=i\int \langle
E_{0}|{\dot{E}}_{0}\rangle $. Therefore the substitution of Eq.~(\ref{adp})
into the instantaneous eigenbasis of $H(t)$, defined by $H(t)|E_{n}(t)%
\rangle =E_{n}(t)|E_{n}(t)\rangle $, yields 
\begin{equation}
H(t)|\psi (t)\rangle =E_{0}(t)|\psi (t)\rangle ,  \label{e0t}
\end{equation}%
which simply states that the wave function is an instantaneous eigenstate of
the Hamiltonian in the adiabatic regime. Substituting Eq.~(\ref{e0t}) into
the Schr\"{o}dinger equation~(\ref{se}) one obtains 
\begin{equation}
i|{\dot{\psi}}\rangle =E_{0}(t)|\psi (t)\rangle .  \label{ase}
\end{equation}%
It is important to observe that the above equation has been derived by using
the fact that \emph{the adiabatic solution must be an instantaneous
eigenstate of the Hamiltonian}. Moreover one can see that the adiabatic wave
function is really a solution of the adiabatic Schr\"{o}dinger equation by
substituting Eq.~(\ref{adp}) into Eq.~(\ref{ase}), from which one obtains 
\begin{equation}
\left( 1-|E_{0}(t)\rangle \langle E_{0}(t)|\right) |{\dot{E}}_{0}(t)\rangle
=0.  \label{eproj}
\end{equation}%
In order to show that Eq.~(\ref{eproj}) is obeyed in the adiabatic regime we
can project this equation by multiplying it by each instantaneous basis
vector $\langle E_{n}(t)|$: 
\begin{equation}
\langle E_{n}(t)|\left( 1-|E_{0}(t)\rangle \langle E_{0}(t)|\right) |{\dot{E}%
}_{0}(t)\rangle =0.
\end{equation}%
Therefore we obtain that the above equation is satisfied, for each $n$, if
the adiabatic constraints [see Eq.~(\ref{vcc})] are obeyed 
\begin{equation}
\left\vert \frac{\langle E_{n}(t)|{\dot{E}}_{0}(t)\rangle }{E_{0}(t)-E_{n}(t)%
}\right\vert \ll 1,\,\,\,(n\neq 0)
\end{equation}


\subsection{The inconsistent step}


Now suppose that we wish to solve the adiabatic Schr\"{o}dinger equation~(%
\ref{ase}), but (incorrectly) ignore the fact that it has been derived by
assuming that $|\psi (t)\rangle $ is an instantaneous eigenstate of $H(t)$.
Then, imposing the initial condition $|\psi (t_{0})\rangle
=|E_{0}(t_{0})\rangle $, one easily finds that Eq.~(\ref{ase}) is satisfied
by the following wave function: 
\begin{equation}
|\psi (t)\rangle =e^{-i\int^{t}E_{0}}|E_{0}(t_{0})\rangle .  \label{fsol}
\end{equation}%
However, this $|\psi (t)\rangle $ is clearly inconsistent with Eq.~(\ref{e0t}%
) and therefore is an illegal solution, since it generally is not an
instantaneous eigenstate of the Hamiltonian. Indeed, the general adiabatic
solution is given by Eq.~(\ref{adp}), which includes the Berry's phase and $%
|E_{0}(t)\rangle $, as opposed to $|E_{0}(t_{0})\rangle $. In fact, if we
take Eq.~(\ref{fsol}) as the adiabatic wave function and substitute it into
Eq.~(\ref{adp}) we obtain 
\begin{equation}
e^{i\beta _{0}}|E_{0}(t)\rangle =|E_{0}(t_{0})\rangle .  \label{inc1}
\end{equation}%
Then multiplying Eq.~(\ref{inc1}) by $\langle E_{0}(t_{0})|$ 
\begin{equation}
e^{i\beta _{0}}\langle E_{0}(t_{0})|E_{0}(t)\rangle =1.  \label{inc2}
\end{equation}%
This inconsistency is precisely the one claimed by Marzlin and Sanders in
Ref. \cite{Marzlin:04}, Eq.~(6). Note that this result is obtained there in
a somewhat more complicated manner, by considering the adiabatic solution of
a time-reversed wave function $|{\bar{\psi}}(t)\rangle =U^{\dagger
}(t,t_{0})|E_{0}(t_{0})\rangle $. We note that the solution for $|{\bar{\psi}%
}(t)\rangle $ [their Eq.~(4)] is very similar to our Eq.~(\ref{fsol}).
Hence, in the same way that we have been led to an inconsistent result due
to a deliberately wrong adiabatic solution for $|\psi (t)\rangle $, Ref. 
\cite{Marzlin:04} has been led to an inconsistent solution for their $|{\bar{%
\psi}}(t)\rangle $.


\section{Conclusion}


We have reviewed the adiabatic dynamics of both closed and open quantum
systems. In the case of closed systems the adiabatic limit is the case where
initial Schr\"{o}dinger-eigenspaces evolve independently, without any
transitions between eigenspaces; this limit can be relaxed and a
perturbation theory can be developed in the number of transitions. In the
case of open systems the notion of Schr\"{o}dinger-eigenspaces is replaced
by independently evolving Lindblad-Jordan blocks. A corresponding
perturbation theory has not yet been developed. We have also shown that the
inconsistency in the adiabatic theorem claimed in Ref.~\cite{Marzlin:04} is
a consequence of an improper adiabatic solution for the wave function. One
arrives at an inconsistent result by taking the \emph{instantaneous}
adiabatic eigenstates and integrating them over \emph{all time} using the
adiabatic Schr\"{o}dinger equation. The adiabatic theorem remains a valuable
and consistent tool for studying the dynamics of slowly evolving quantum
systems.


\section*{Acknowledgments}


The research of the authors is sponsored by CNPq-Brazil (to M.S.S.), and the
Sloan Foundation, PREA and NSERC (to D.A.L.).


\begin{thebibliography}{99}
\bibitem{Ehrenfest:16} P. Ehrenfest, {Ann. d. Phys.} \textbf{51}, 327 (1916).

\bibitem{Born:28} {M. Born and V. Fock}, {Zeit. f. Physik} \textbf{51}, 165
(1928).

\bibitem{Kato:50} {T. Kato}, {J. Phys. Soc. Jap.} \textbf{5}, 435 (1950).

\bibitem{Messiah:book} {A. Messiah}, \emph{Quantum mechanics} ({North-Holland%
}, {Amsterdam}, 1962), Vol.~2.

\bibitem{Berry:87} M.V. Berry, {Proc. R. Soc. London} A \textbf{414}, 31
(1987).

\bibitem{Nakagawa:87} N. Nakagawa, {Ann. Phys.} \textbf{179}, 145 (1987).

\bibitem{Sun:88} C.P. Sun, {J. Phys. A} \textbf{21}, 1595 (1988).

\bibitem{Wu:89} Z. Wu, {Phys. Rev. A} \textbf{40}, 2184 (1989).

\bibitem{Landau:32} {L.D. Landau}, Zeitschrift \textbf{2}, 46 (1932).

\bibitem{Zener:32} {C. Zener}, Proc. Roy. Soc. London Ser. A \textbf{137},
696 (1932).

\bibitem{Gellmann:51} {M. Gell-Mann and F. Low}, Phys. Rev. \textbf{84}, 350
(1951).

\bibitem{Berry:84} {M.V. Berry}, Proc. Roy. Soc. (Lond.) \textbf{392}, 45
(1989).

\bibitem{Wilczek:84} {F. Wilczek and A. Zee}, Phys. Rev. Lett. \textbf{52},
2111 (1984).

\bibitem{ZanardiRasseti:99} {P. Zanardi and M. Rasetti}, Phys. Lett. A 
\textbf{264}, 94 (1999).

\bibitem{ZanardiRasseti:2000} {J. Pachos, P. Zanardi, and M. Rasetti}, Phys.
Rev. A \textbf{61}, 010305 (2000).

\bibitem{Ekert-Nature} {J.A. Jones, V. Vedral, A. Ekert, and G. Castagnoli},
Nature \textbf{403}, 869 (2000).

\bibitem{Pachos:00} {J. Pachos and S. Chountasis}, Phys. Rev. A \textbf{62},
052318 (2000).

\bibitem{Duan-Science:01} {L.-M. Duan, J. I. Cirac, and P. Zoller}, Science 
\textbf{292}, 1695 (2001).

\bibitem{Pachos:02} {I. Fuentes-Guridi, J. Pachos, S. Bose, V. Vedral, and
S. Choi}, Phys. Rev. A \textbf{66}, 022102 (2002).

\bibitem{Fazio:03} {L. Faoro, J. Siewert, and R. Fazio}, Phys. Rev. Lett. 
\textbf{90}, 028301 (2003).

\bibitem{Farhi:00} {E. Farhi, J. Goldstone, S. Gutmann, and M. Sipser}, LANL
Preprint quant-ph/0001106.

\bibitem{Farhi:01} {E. Farhi, J. Goldstone, S. Gutmann, J. Lapan, A.
Lundgren, and D. Preda}, Science \textbf{292}, 472 (2001).

\bibitem{Aharonov:04} D. Aharonov, W. v. Dam, J. Kempe, Z. Landau, S. Lloyd,
O. Regev, LANL Preprint quant-ph/0405098.

\bibitem{SarandyLidar:04} {M.S. Sarandy and D.A. Lidar}, LANL Preprint
quant-ph/0404147, Phys. Rev. A (2004), in press.

\bibitem{LidarWhaley:03} {D.A. Lidar and K.B. Whaley}, in \emph{Irreversible
Quantum Dynamics}, Vol.~622 of \emph{Lecture Notes in Physics}, edited by {%
F. Benatti and R. Floreanini} ({Springer}, {Berlin}, 2003), p. 83, LANL
Preprint quant-ph/0301032.

\bibitem{Marzlin:04} {K.-P. Marzlin and B.C. Sanders},  
Phys. Rev. Lett. {\bf 93}, 160408 (2004).

\bibitem{Tong:04} {D.M. Tong, K. Singh, L.C. Kwek, and C.H. Oh}, LANL
Preprint quant-ph/0406163 (2004).

\bibitem{Pati:04} {A.K. Pati and A.K. Rajagopal}, LANL Preprint
quant-ph/0405129 (2004).

\bibitem{Wu:04} Z. Wu and H. Yang, LANL Preprint quant-ph/0410118 (2004).

\bibitem{Mostafazadeh:book} {A. Mostafazadeh}, \emph{Dynamical Invariants,
Adiabatic Approximation, and the Geometric Phase} ({Nova Science Publishers}%
, {New York}, 2001).

\bibitem{Avron:98} J.E. Avron and A. Elgart, Phys. Rev. A {\bf 58}, 4300 (1998); 
Commun. Math. Phys. {\bf 203}, 445 (1999).

\bibitem{Gottfried:book} {K. Gottfried and T.-M. Yan}, \emph{Quantum
Mechanics: Fundamentals} ({Springer}, {New York}, 2003).

\bibitem{Churchill:book} {J.W. Brown and R.V. Churchill}, \emph{Fourier
series and boundary value problems} ({McGraw-Hill}, {New York}, 1993).

\bibitem{RiemannLebesgue} The Riemann-Lebesgue lemma can be stated through
the proposition: Let $f:[a,b]\rightarrow \mathbf{C}$ be an integrable
function on the interval $[a,b]$. Then $\int_{a}^{b}\,dx\,e^{inx}f(x)%
\rightarrow 0$ as $n\rightarrow \pm \infty $.

\bibitem{geomphase-open} L.-B. Fu, J.-L. Chen, J. Phys. A \textbf{37}, 3699
(2004); E. Sj\"{o}qvist, LANL Preprint quant-ph/0404174; K.-P. Marzlin, S.
Ghose, B.C. Sanders, LANL Preprint quant-ph/0405052; D.M. Tong, E. Sj\"{o}%
qvist, L.C. Kwek, C.H. Oh, LANL Preprint quant-ph/0405092; R.S. Whitney, Y.
Makhlin, A. Shnirman, Y. Gefen, LANL Preprint cond-mat/0405267;\ I.
Kamleitner, J.D. Cresser, B.C. Sanders,\ LANL Preprint quant-ph/0406018.

\bibitem{Breuer:book} {H.-P. Breuer and F. Petruccione}, \emph{The Theory of
Open Quantum Systems} ({Oxford University Press}, Oxford, 2002).

\bibitem{Kraus:71} {K. Kraus}, Ann. of Phys. \textbf{64}, 311 (1971).

\bibitem{Alicki:87} {R. Alicki and K. Lendi}, \emph{Quantum Dynamical
Semigroups and Applications}, No.~286 in \emph{Lecture Notes in Physics} ({%
Springer-Verlag}, Berlin, 1987).

\bibitem{Gorini:76} {V. Gorini, A. Kossakowski, and E.C.G Sudarshan}, J.
Math. Phys. \textbf{17}, 821 (1976).

\bibitem{Lindblad:76} {G. Lindblad}, Commun. Math. Phys. \textbf{48}, 119
(1976).

\bibitem{Lidar:CP01} {D.A. Lidar, Z. Bihary, and K.B. Whaley}, Chem. Phys. 
\textbf{268}, 35 (2001).

\bibitem{Lendi:86} K. Lendi, Phys. Rev. A {\bf 33}, 3358 (1986).

\bibitem{Breuer:04} H.-P. Breuer, Phys. Rev. A {\bf 70}, 012106 (2004).

\bibitem{Horn:book} {R.A. Horn and C.R. Johnson}, \emph{Matrix Analysis} ({%
Cambridge University Press}, {Cambridge, UK}, 1999).

\bibitem{Thomaz:03}
{A.C.A. Pinto and M.T. Thomaz}, J. Phys. A {\bf 36},  7461  (2003).

\bibitem{Carollo:04}
{A. Carollo, I. Fuentes-Guridi, M.F. Santos, and V. Vedral}, Phys. Rev. Lett.
  {\bf 92},  020402  (2004).

\bibitem{Sanders:04} I. Kamleitner, J.D. Cresser, and B.C. Sanders, e-print quant-ph/0406018 (2004).

\end{thebibliography}

\end{document}